\documentclass[preprint,showpacs,preprintnumbers,amsmath,amssymb,prl]{revtex4}

\usepackage{graphicx}
\usepackage{dcolumn}
\usepackage{bm}
\usepackage{doublespace}

\begin{document}


\title{Microscopic theory of network glasses}

\author{Randall W. Hall}
\affiliation{Department of Chemistry \\ Louisiana State
University\\Baton Rouge, La. 70803-1804} 

\author{Peter G. Wolynes}%
\affiliation{Department of Chemistry and Biochemistry\\University of California, San Diego\\La Jolla, Ca. 92093-0332}%

\date{\today}

\begin{abstract}
A molecular theory of the glass transition of network forming
liquids is developed using a combination of self-consistent phonon
and liquid state approaches.  Both the dynamical transition and
the entropy crisis characteristic of random first order
transitions are mapped out as a function of the degree of bonding
and the density. Using a scaling relation for a soft-core model to
crudely translate the densities into temperatures, the theory
predicts that the ratio of the dynamical transition temperature to
the laboratory transition temperature rises as the degree of
bonding increases, while the Kauzmann temperature falls relative
to the laboratory transition. These results indicate why highly
coordinated liquids should be "strong" while van der Waals liquids
without coordination are "fragile".
\end{abstract}

\maketitle

While much of the phenomenology of the glass transition is
universal, shared by all dense liquids, everyday glasses are
formed from liquids that possess a network structure at the
molecular level. The universal features of the liquid-glass
transition such as the Kauzmann entropy crisis and the
Vogel-Fulcher rate law have been explained using theories based on
an underlying random first-order
transition\cite{Wolynes1992,Thirumalai1995,Wolynes2000}, like that
of mean field models with one step replica symmetry
breaking\cite{Sompolinksky1985}. Furthermore, the variations of
the mean relaxation time\cite{Wolynes2000} and degree of
nonexponentiality\cite{Wolynes2001A} of relaxation from substance
to substance have been quantitatively predicted using this
approach, provided the configurational heat capacity is known as
input, thus quantitatively explaining the patterns of behavior
verbally characterized using the expression "liquid
fragility"\cite{Angell1985}. Detailed microscopic calculations
buttressing these developments are based on a model of  a dense
interacting atomic fluid in which the particles interact
  via van der Waals forces alone.
Is this starting point appropriate for the network glasses?  To
address this issue, we explore the glass transition in a model of
a network forming liquid. The calculations presented are similar
in spirit to those made earlier for atomic fluids. Our model
assumes cross-linking in a liquid is fluctuating, not quenched as
in models of rubber\cite{Goldenfeld1989}, taking the mean degree
of crosslinking as given from experiment or from a separate
calculation explicitly accounting for the equilibria between the
various different bonded species. These calculations suggest a
connection between the glassy behavior of network systems as
viewed by random first order transition theory and rigidity
percolation. Rigidity percolation features prominently in
Phillips's discussion of the role of bonding constraints in glass
formation\cite{Phillips1985}, as well as in the extensive studies
of Thorpe\cite{Thorpe1985A,Thorpe1999}. A major result of the
present calculations is that the density of non-bonded
interactions needed to trigger energy landscape dominated glassy
behavior decreases when bonding increases.  It follows that,
consistent with experiment, highly cross-linked network glasses
should show activated dynamics even at very high temperatures. The
theory still predicts a crossover to non-activated dynamics for
sufficiently high temperature as has recently been inferred in
computer simulations of silica\cite{Kob2001,Verrocchio2002}.

The self-consistent phonon method applied to the amorphous state
provides our starting point\cite{Fixman1969,Wolynes1984}. This
scheme is closely connected to approaches based on density
functionals\cite{Wolynes1985}. It has recently been justified by
an elegant clone or replica approach\cite{Peliti1999,Mezard1999}.
We actually use a hybrid of both density functional and
self-consistent phonon calculations. For the hard sphere system,
both approaches agree there is a critical density $\rho_A $, below
which localized density distributions corresponding to amorphous
packings are mechanically unstable to even small thermal
fluctuations. $\rho_A \sigma^3 $  from the self-consistent theory
is approximately 1.0 for hard spheres, where $\sigma$ is the hard
sphere diameter. This is in agreement with the computer simulated
dynamical transition\cite{Woodcock1981}. This instability, from
the vitrified side, coincides with the transition found in
dynamical mode coupling approaches which start from the liquid
side\cite{Wolynes1992,Thirumalai1995,Gotze1999,Gotze1984}. The
thermal vibration allowed at $\rho_A $ in a frozen amorphous state
gives mean square fluctuations $<\delta r^2
> = \alpha^{-1}$, where $\alpha$ is an effective spring constant
and is finite at the transition. $\alpha$ provides an analog of
the Edwards-Anderson order parameter for spins and exhibits
discontinuous replica symmetry breaking. This limit of stability
for an amorphous system is comparable to the Lindemann criterion
that describes melting of crystalline solids. At densities above
 $\rho_A $ many different
mechanically stable amorphous packings or free energy minima can
exist. At densities beyond $\rho_A $, however, the configurational
entropy of these minima decreases until it vanishes at a density
$\rho_K $.  Analogously, for soft-sphere models at fixed densities
one can find a temperature $T_{A}$, signaling  the onset of glassy
dynamics (``landscape determined''\cite{Stillinger1998})and a
still lower temperature $T_{K}$, the Kauzmann temperature where
the configurational
 entropy would vanish.
 Between these densities or temperatures, the dynamics can be described
using the concept of entropic
droplets\cite{Wolynes1992,Wolynes2000,Thirumalai1995,Wolynes2001A}.
The activation free
 energy diverges as $T_{K}$ is approached, consistent with the Vogel-Fulcher law $\Delta F^\dag =DT_K /(T-T_K )$.
A microscopic calculation of the energy cost of forming an
entropic droplet quantitatively explains the experimentally
observed variation of D with the heat capacity drop upon
vitrification, $\Delta C_p $\cite{Wolynes2000}. This correlation
relies on the idea that the Lindemann ratio near $T_K $ should
depend only weakly on the detailed force law. We will check this
supposition for our network glass model.

The free energy is computed for a given amorphous packing as a
function of $\alpha$. This free energy should then be averaged
over packings. Each atom can be assigned to a cell and the density
can be written as
$\rho(\mathbf{r})=\sum_i \rho_i =\sum_i
\left(\frac{\alpha_{i}}{\pi}\right)^{3/2}
  exp\left[-\alpha_{i} (\mathbf{r}-\mathbf{R}_i )^2 \right] \label{rho}$
where the locations $\{\mathbf{R}_i\}$, characterize the packing
as a set of fiducial locations around which the atoms vibrate. The
free energy as a function of $\{\alpha_i \}$, $F(\{\alpha_i \})$,
is analogous to the inter replica potential given as a function of
q for spin glasses\cite{Parisi1997}.

More generally, the thermal vibrations are described by a tensor
$\boldsymbol{\alpha}_i $ defined by
\begin{eqnarray}
\boldsymbol{\alpha}^{-1}_{i}&=& \left<(\mathbf{r}_i -\mathbf{R}_i
) (\mathbf{r}_i -\mathbf{R}_i ) \right>_{W^{eff}_i
}\label{selfconsistency}
\end{eqnarray}
where the averaging is with respect to the effective potential
(dependent on $\{\alpha_j \}$)
\begin{eqnarray}
\exp\left( -\beta W^{eff}_i \right)\equiv\Pi_{i\ne
j}\exp\left(-\beta V^{eff}(|\mathbf{r}_i -\mathbf{R}_j |)\right)
 =
 \Pi_{i\ne j} \int \mathbf{dr}_j \ \rho_{j}(\mathbf{r}_j ) \ \exp\left(
 -\frac{\beta}{2}V(r_{ij})\right)\label{effectivepot}
\end{eqnarray}
 where $V(r)$ is the potential energy. For an isotropic glass under the assumption that all
$\alpha_i $ are the same $V^{eff}$ becomes essentially a function
of a scalar distance R, $V^{eff}(R)$, and one
finds\cite{Wolynes1984}
$ \alpha = \frac{\rho}{6}\int dR\ \ g(R)\ \ \nabla^2 \left(\beta
V^{eff}(R)\right)\ $
 and a free energy
$ \frac{F(\alpha)}{Nk_B T}=f^{ideal} + \rho \ \ \int dR\ \ g(R)\ \
V^{eff}(R)$.
Although a good approximation for large $\alpha$, the
self-consistent phonon theory is not accurate for low $\alpha$. At
low $\alpha$, an expansion around the uniform fluid can be used,
but this depends on knowing the direct correlation function which
is not available
 for network fluids.  When no bonds are present, alternatively one can use the functional first suggested
by Tarazona\cite{Tarazona1984A} and generalized by
Ashcroft\cite{Ashcroft1999}. This functional takes the form
$ \frac{F}{Nk_b T}\equiv f=\frac{1}{N}\int dr \rho(r) \left[ \log
(\rho(r)) -1 \right]+ \Psi(\eta)$
where $\Psi$ is the interaction part of the free energy for a
monatomic liquid or glass (which we take to be the
Carnahan-Starling expression) and $\eta$ is the packing fraction
($\pi\rho\sigma^{3}/6$). For the network forming liquids we will
correct this free energy using a modification of the result of
Erukhimovich\cite{Erukhimovich2001}, which provides for the
entropy loss of forming bonds.

For the network forming liquid and glass, we separate the
Hamiltonian into  bonding and non-bonding parts $V=\sum_{i,j\
bonded}V_{b}(r_{ij}) + \sum_{i,j\ non-bonded}V_{nb}(r_{ij}) $.
Each pair i and j is considered to be either bonded on non-bonded
and thus have an interaction described by either $V_{b}$ or
$V_{nb}$. The molecular units which are considered to be bonded
have a locally harmonic potential energy $V_b (r_{ij}) =V_0
+\frac{1}{2}\gamma^b (r_{ij}-d_b )^2$ with $d_b $ the equilibrium
bond distance. $V_{nb} $, on the other hand, represents non-bonded
interactions and therefore is taken to be a hard or soft sphere.
Notice this potential does not explicitly describe, for any given
configuration, whether a specific given pair should be considered
bonded or not. This requires writing the potential using manybody
switching functions\cite{Stillinger1985,Rossky1992}. These
manybody functions would reflect the saturation of chemical
bonding tendencies (and would be needed for simulations.) For such
a model the theory of associated liquids yields a (renormalized)
equilibrium constant for the bonding reactions in terms of the
underlying potential\cite{Andersen1983}. We will take the
 fraction of bonded pairs to be given and to be temperature and density independent. Of course,
density and temperature dependent changes of the association
equilibrium would be easy to include, but would distract from
understanding how the glassy physics itself depends on
temperature. Provided the association equilibrium constants are
not paradoxically related to the density (i.e., decrease partially
with $\rho$) the trends in any event will be as shown.

For the network system, the self-consistent phonon calculation
must distinguish the two types of interactions. For specificity we
assume any given atom can form 0-4 bonds, in a tetrahedral
pattern. Other spatial patterns are easily treated as well. At
large $\alpha$ the effective potential coming from a bonded pair
is essentially
$V_{b}^{eff}(\mathbf{R})=\frac{V_0 }{2}+\frac{\gamma^b
}{4}\left(|\mathbf{R}|-d_b \right)^2 $
where the factor of 1/2 comes from the self-consistent
formulation.  We set the bond distance $d_b $ equal to the hard
sphere diameter for the non-bonded atoms and choose a zero of
energy such that $V_0 = 0$. In numerical calculations we used
$\beta\gamma^b = 300$, approximately the value appropriate for
silica potentials\cite{vanSanten1990}. The probability that a
given atom has n-bonds, $p_n $ is given by
Erukhimovich\cite{Erukhimovich2001}
$ p_n =\binom{4}{n}\Gamma^{n}(1-\Gamma)^{4-n} $
with $\Gamma = \frac{n_b }{4}$ and $n_b $ is the average number of
bonds per atom.

In our calculations the basic pair distribution for non-bonded
interactions, $g(R)$ is taken to be the Percus-Yevick radial
distribution function modified for high density in the standard
way. To account for the presence of bonded interactions that block
the approach of other partners, $g(R)$ for an atom with $n$ bonds
(denoted $g_{n}(R) $) is scaled by
$\max\left(0,\frac{<nn>_{\rho}-n}{<nn>_{\rho}}\right)$ for
distances less than the first minimum of g(R). $<nn>_{\rho}$ is
the number of nearest neighbors at total density $\rho $, obtained
by integrating the first peak of g(R).

To each of the 5 types of atoms, those with 0, 1, 2, 3, or 4
bonds, we assign a mean $\alpha_n $. In this paper we use the
\textit{uncoupled oscillator} implementation of the
self-consistent phonon theory\cite{Wolynes1984} to determine the
free energy. Essentially, this approximation treats the glass as a
collection of Einstein oscillators. The derivation in Stoessel and
Wolynes's early paper is lengthy in detail but when applied to the
present potential one obtains
%

\begin{eqnarray}
F_{scp}/Nk_B T=\sum_{n=0}^{4}p_n \left[V^{eff}_{b,n}(d_b
)+\rho\int dR
g_n (R) V^{eff}_{nb}(R)\right] \nonumber \\
-\ln\int_{v}\ dr^{N} e^{-\sum_{n=0}^{4}\frac{p_n
}{2}r_{i}\cdot\left[\nabla\nabla\beta
V^{eff}_{b,n}(R_{i})+\mathbf{I}\alpha^{nb}_{n}\right]\cdot r_{i}}
\label{qnet}
\end{eqnarray}
where $V^{eff}_{b,n}$ is the sum of $V^{eff}_{b}$ for an atom with
$n$ bonds over its bonds (and includes correlations between bonds
through the use of a tetrahedral lattice), $V^{eff}_{nb}$ is
defined by Eqn.~\ref{effectivepot} using only the non-bonding
potential energy and \textbf{$g_n $}, and $\alpha^{nb}_n $ is the
non-bonding contribution to $\alpha_n $. With a proper choice of
local axes for each $r_i $, the argument of the exponential can be
diagonalized, leading to 3 principal values of the vibrational
tensor $\alpha_{1,n}$, $\alpha_{2,n}$, and $\alpha_{3,n}$.
For the free energy to be consistent with
Eqn.~\ref{selfconsistency}, we must have $\alpha_n$  satisfy
$3\alpha_{n}^{-1}=\alpha^{-1}_{1,n}  +\alpha^{-1}_{2,n}  +
\alpha^{-1}_{3,n}\label{scf_net}$
 Note that for a glass transition to occur, all 3 components of
$\alpha_n $ must be large. Given a bonding pattern, the
$\{\alpha\}$ can be determined and, hence, the free energy.

$F_{scp}$ is the free energy of the large $\alpha$ glassy minimum.
We compute the free energy of the network liquid with $\alpha = 0$
using the Carnahan-Starling expression for a non-bonded liquid
plus a modification of Erukhimovich's result for equilibrium
network materials\cite{Erukhimovich2001}:
$F_{liq}/Nk_B T=\Psi(\eta)+4\left[\Gamma\ln\Gamma +
(1-\Gamma)\ln(1-\Gamma)\right]+\ln\left(\frac{\rho}{e}\right)-
\frac{n_b
 }{2}\ln\left(\frac{n_b \rho}{e}\right)+\frac{F_{scp}^{b}}{Nk_{B}T}
 \label{fliq}$
where $F_{scp}^{b}$ is obtained from Eqn.~\ref{qnet} with $\alpha
= 0$ and only the bonding terms. It corresponds to the entropy
loss from bonding computed by Erukhimovich\cite{Erukhimovich2001}.

The self-consistent phonon theory alone allows us to find $\rho_A
$, as the lowest density giving a non-zero $\alpha $, which thus
corresponds to the onset of activated behavior while $\rho_K $,
the "Kauzmann" density at which the free energies of the liquid
and glass match and the configurational entropy vanishes also
relies on the bonding correction to $F(\alpha=0)$. The ratio
$\rho_A / \rho_K $ dimensionlessly characterizes the thermodynamic
aspects completely. To compare to the kinetic laboratory glass
transition, we note the laboratory glass transition defined to
occur when the viscosity reaches $10^{14}$ p. Previous work using
random first order theory predicts this to occur when the
configurational entropy is about 1.0 $k_B $. To make the
translation of our thermodynamic results to the laboratory
transition density in what follows we will therefore mean by
$\rho_G $ the density where the liquid and glass free energies
differ by  1.0 $k_B T$ per particle. The universality of
configurational entropy at the laboratory transition is well
confirmed, also by experiment.

\begin{figure}
\centerline{\includegraphics*[scale=.3,keepaspectratio=true]{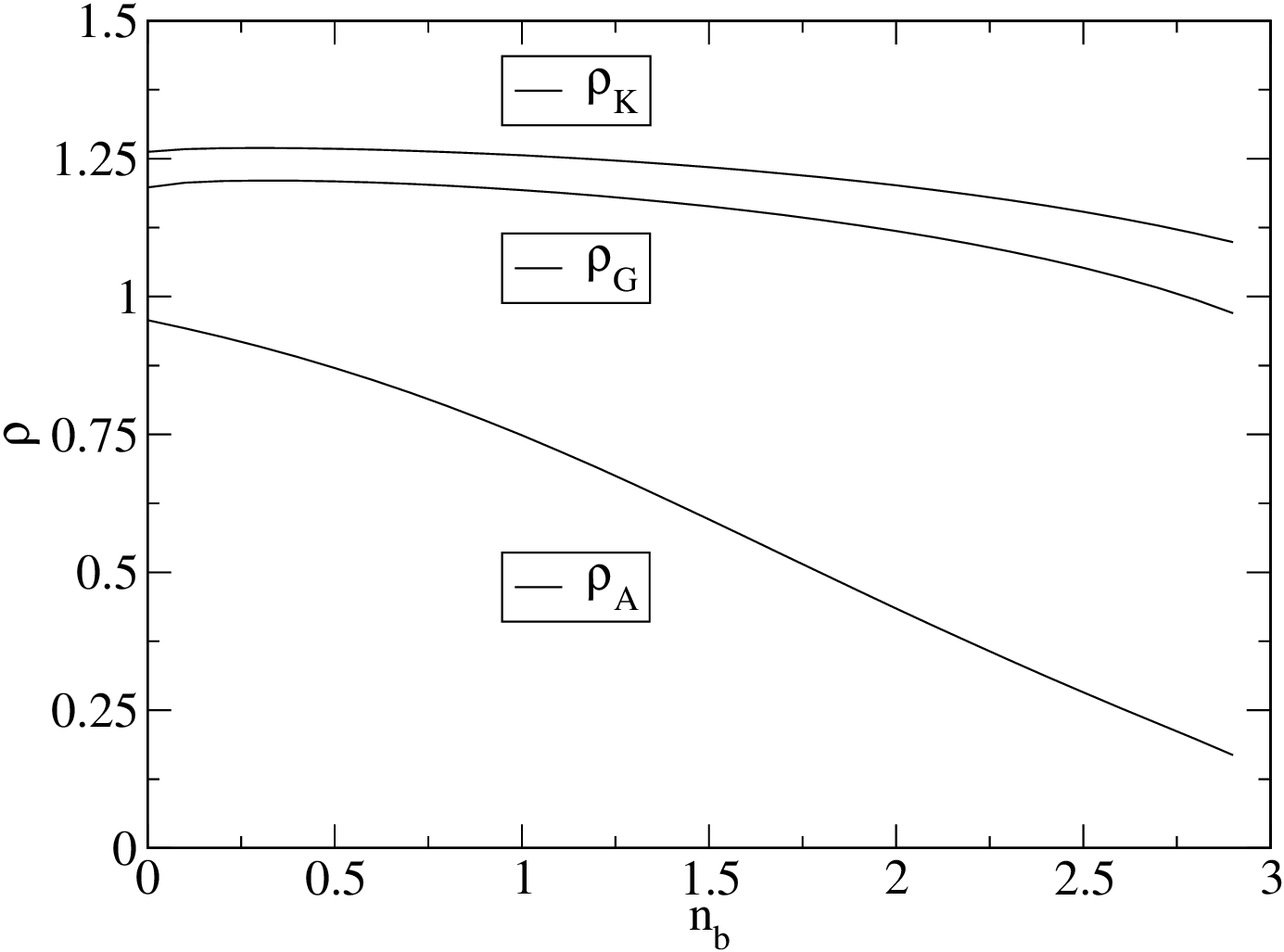}}
\caption{\footnotesize Plot of $\rho_A $, $\rho_K $, and $\rho_G $
versus $n_b  $. \label{rhoplot}}
\end{figure}

\begin{figure}
\centerline{\includegraphics*[scale=.3,keepaspectratio=true]{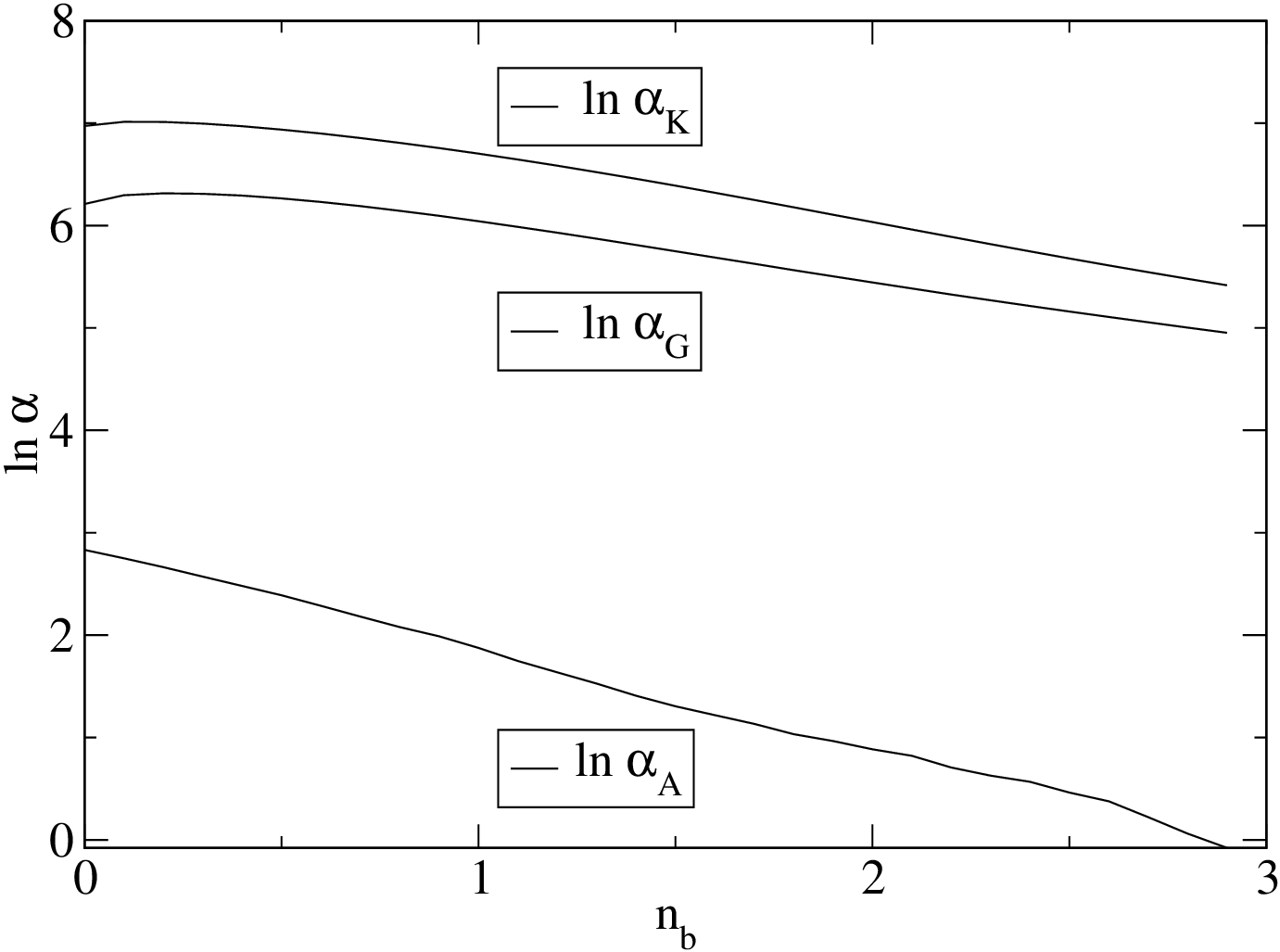}}
\caption{\footnotesize Plot of $\ln\alpha_A $, $\ln\alpha_K $, and
$\ln\alpha_G $ versus $n_b $. \label{alphaplot}}
\end{figure}

$\rho_A $, $\rho_K $, and $\rho_G $ as functions of the average
number of bonds $n_b $ are shown in Fig.~\ref{rhoplot}. For $n_b
> 3$, $\rho_A $ vanishes, indicating that the rigidity percolation
occurs when $n_b  = 3$ in our model. For comparison, Thorpe more
precisely defining  rigidity percolation as the absence of zero
mode vibrational frequencies, finds $n_b $ = 2.4. For $n_b
> 3$, even at low densities the network is a rubber that can still vitrify at a higher densities.
Fig.~\ref{alphaplot} shows the effective spring constants at these
densities. For the hard sphere model\cite{Wolynes1984} the
uncoupled oscillator approximation leads to a somewhat lower
$\rho_A $ than simulation and to relatively small, but
nevertheless finite, values of $\alpha_A $. Quantitatively, the
more complete coupled oscillator approximation (which uses a chain
summation to treat the glass as a collection of Debye
oscillators\cite{Wolynes1984}) leads to a larger $\rho_A $ in
better agreement with computer studies. So we would expect
quantitative modification by taking better account of the coupling
but expect the same trend, namely that as the amount of bonding is
increased, the transition densities decrease and therefore the
transition temperatures increase. This theory then explains why
the addition of a non-bonding impurity into a networked glass will
lower the glass transition temperature, as known since ancient
times for silica\cite{Pliny}. In physical terms, the non-bonding
transition densities decrease because, as seen in Eqn.~\ref{qnet},
as the contribution from bonding to the vibrational force
increases, the remaining contribution needed from non-bonding
interactions for self-consistency decreases.

\begin{figure}
\centerline{\includegraphics*[scale=.3,keepaspectratio=true]{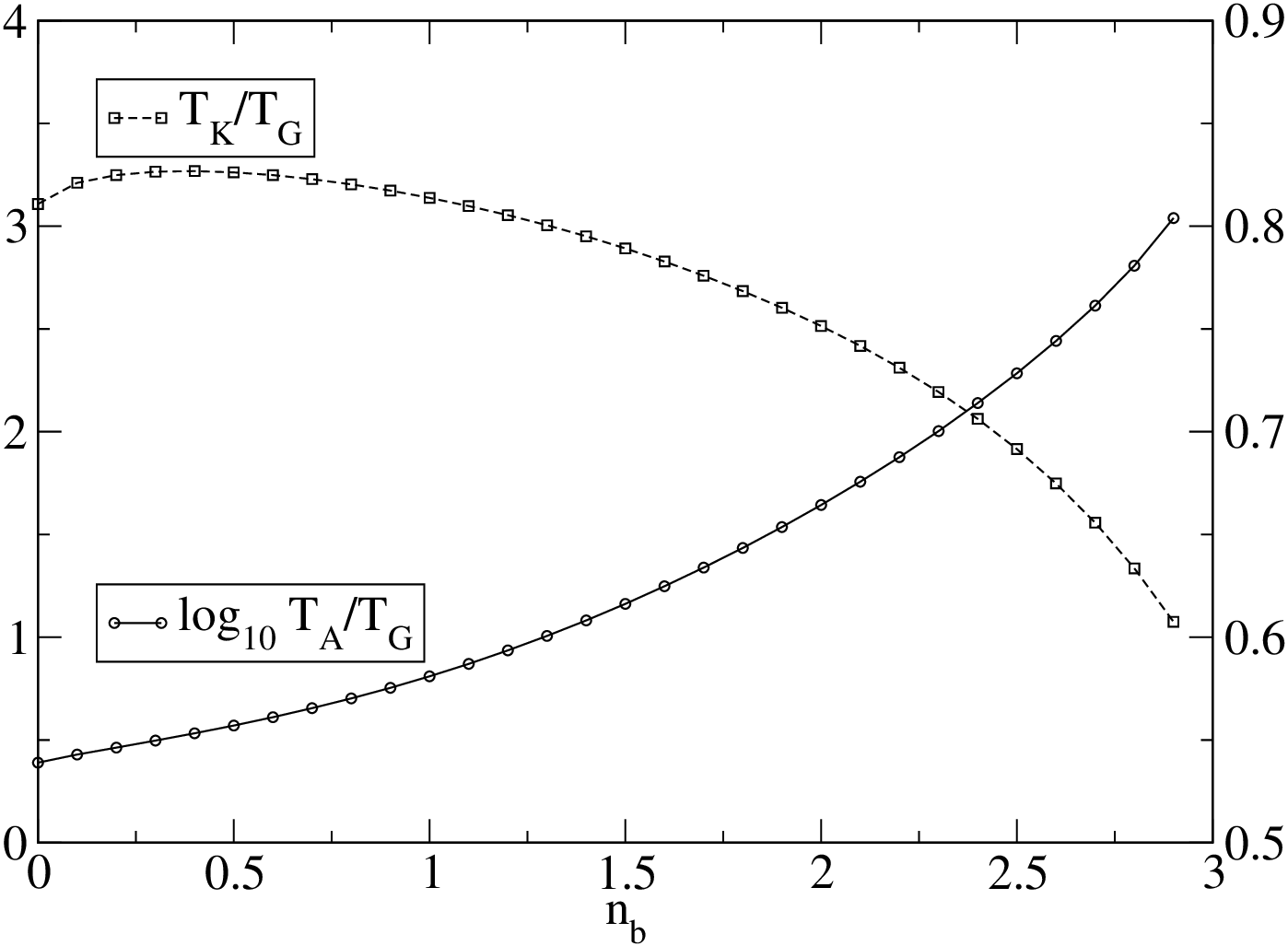}}
\caption{\footnotesize Plot of $\log_{10}(T_A /T_G )$ (scale shown
on left axis)  and $T_K /T_G $ (scale shown on right axis) versus
$n_b $. \label{Tplot}}
\end{figure}

It is clear that $\rho_A $ has a much sharper dependence on
density than either $\rho_K $ or $\rho_G $. The degree of bonding
has a very direct effect on the vibrational force constant, being
proportional to the average Laplacian of the effective potential,
but only an indirect effect on entropy, entering through the more
slowly varying (in space) Boltzmann factor. The theory explains
the change of fragility with bonding. To see this we note that
real liquids have both repulsive cores of finite strength and
attractions which conspire in setting the equation of state. Yet
assuming the attractions provide merely a smooth background
potential the density dependences of this model can be translated
into temperature dependences. For a soft potential $V\sim r^{-n}$
we have the scaling relation $T_1 /T_2  = \left(\rho_{2}/\rho_1
\right)^{3/n}$. Using this relation for $n = 12 $ yields  $T_A
/T_G $ and $T_K / T_G $ as plotted in Fig.~\ref{Tplot}.
Experimental measurements of $T_K / T_G $ give 0.82 for the
fragile glass orthoterphenyl\cite{Angell1997}, while for the
strong glass silica the ratio has recently been
determined\cite{Sipp2001} to be 0.36. The uncoupled oscillator
approximation results in a variation from 0.82 for the
non-networked glass to 0.6 at the approximate rigidity percolation
limit, a variation somewhat smaller in magnitude. Doubtless, the
crudeness of this result is due to our use of a scaling
approximation appropriate to a simple, purely repulsive model and
better agreement with experiment would be obtained by including
attractive forces in our estimates. We see that $T_A /T_G $
increases dramatically with $n_b $, while $T_K /T_G $ decreases
modestly. Thus highly bonded liquids have large temperature ranges
of nearly Arrhenius activated behavior while liquids with few
crosslinks have smaller ranges of strongly non-Arrhenius activated
behavior.

For the uncoupled oscillator treatment $\alpha_A $ falls rapidly
with density to very small values as crosslinking increases. The
universality of the Lindemann ratio previously assumed by Wolynes
and coworkers\cite{Wolynes2000} would seem in doubt. Yet this is
not so. First, we know the coupled oscillator approximation will
give a larger $\rho_A $ therefore decreasing the range of
variation of $\alpha_A $. For $n_b
 = 0$, the coupled oscillator gives a value of
$\alpha_A \approx 100$, while the uncoupled theory gives $\alpha_A
\approx 20$\cite{Wolynes1984}. More important, even at the
uncoupled level $\alpha_K $ and $\alpha_G $ are roughly constant
with increasing bonding. A careful examination of the earlier
analysis shows that these are the appropriate Lindemann-like
ratios to determine the activation barriers in the strongly glassy
regime. The predicted universal relation between D and $\Delta C_p
$, relying on a universal $\alpha_G $, remains therefore sound.

The combination of self-consistent phonon and liquid state
approaches provides a microscopic picture of network glasses that
explains the dependence of the characteristic glass densities and,
less directly, temperatures on degree of crosslinking and this
results  agrees with experiment. Doubtless, many of the detailed
liquid state approximations made here can be improved but it is
unlikely such modifications will change the qualitative story. In
particular, the attractive forces that go beyond the soft-sphere
model should probably be included to achieve quantitative
agreement with experiment. This model combined with a theory of
the association equilibria in forming the network provides a first
principles treatment of glass properties starting from the
underlying intermolecular forces.

RWH acknowledges NSF Grant. No. 9977124 for support.

\bibliography{networkglass}
\end{document}